\documentclass{article}
\usepackage{spconf}
\usepackage{cite}
\usepackage{amsmath,amssymb,amsfonts}
\usepackage{algorithmic}
\usepackage{graphicx}
\usepackage{textcomp}
\usepackage{xcolor}
\usepackage{microtype}                 
\PassOptionsToPackage{warn}{textcomp}  
\usepackage{textcomp}                  
\usepackage{mathptmx}                  
\usepackage{times}                     
\usepackage{cite}                      
\usepackage{tabu}                      
\usepackage{booktabs}                  
\usepackage{multirow}
\usepackage{url}
\usepackage{amssymb}
\usepackage{amsmath}
\usepackage{float}
\usepackage{wrapfig}
\usepackage{subfigure}
\usepackage{color}
\usepackage{bm}                        
\usepackage{mathrsfs}                  
\usepackage{amsmath}                   
\usepackage[cmintegrals]{newtxmath}    
\usepackage{paralist}
\usepackage{hyperref}
\usepackage{stfloats}
\usepackage[ruled,vlined]{algorithm2e}

\newcommand{\agent}[1]{\textbf{#1}}

\graphicspath{{figs/}{figures/}{pictures/}{images/}{./}} 

\title{ArchAgent: Scalable Legacy Software Architecture Recovery with LLMs}
\name{Rusheng Pan$^{\star \dagger}$ \qquad Bingcheng Mao$^{\star}$ \qquad Tianyi Ma$^{\star}$ \qquad Zhenhua Ling$^{\dagger}$}
\address{$^{\star}$HiThink Research, Hanzhou, China\\
      $^{\dagger}$University of Science and Technology of China, Hefei, China}

\begin{document}
%
\maketitle
\begin{abstract}
Recovering accurate architecture from large-scale legacy software is hindered by architectural drift, missing relations, and the limited context of Large Language Models (LLMs). We present ArchAgent, a scalable agent-based framework that combines static analysis, adaptive code segmentation, and LLM-powered synthesis to reconstruct multiview, business-aligned architectures from cross-repository codebases. ArchAgent introduces scalable diagram generation with contextual pruning and integrates cross-repository data to identify business-critical modules. Evaluations of typical large-scale GitHub projects show significant improvements over existing benchmarks. An ablation study confirms that dependency context improves the accuracy of generated architectures of production-level repositories, and a real-world case study demonstrates effective recovery of critical business logics from legacy projects. The dataset is available at \url{https://github.com/panrusheng/arch-eval-benchmark}. 
\end{abstract}
\begin{keywords}
Software architecture recovery, code repository, cross-repository context, large language models
\end{keywords}

\section{Introduction}\label{sec:introduction}

Software architecture provides foundational abstractions for designing and maintaining complex systems by defining high-level structures, component interrelationships, and key quality attributes such as scalability and maintainability~\cite{perry1992foundations}. However, in large-scale industrial systems, architectural drift often occurs due to requirement changes and ad hoc modifications, leading to poorly documented and difficult-to-recover implementations~\cite{DBLP:conf/fase/Dayani-FardYMA05}.

Recent advances in Large Language Models (LLMs) have enabled automation in code-related tasks like generation and vulnerability detection~\cite{chen2021evaluating}, motivating their use in architectural recovery from source code~\cite{li2022competition, esposito2025generative}. Yet, LLM-based approaches remain limited for repository-level architecture reconstruction due to constrained context windows, sensitivity to code complexity, and lack of integrated multiview perspectives~\cite{DBLP:journals/infsof/NguyenVVN25, dhar2024leveraging}. These limitations are particularly evident in two key areas. \textbf{1. Context limitations.} Industrial codebases often exceed millions of tokens, far exceeding the context window of modern LLMs ($\sim$200k tokens)~\cite{oskooei2025repository}. Although retrieval techniques and tools (e.g. Retrieval Augmented Generation and Model Context Protocal) help with localized understanding~\cite{li2024retrieval}, they do not support coherent system-level architectural synthesis. \textbf{2. Missing business semantics.} Legacy systems typically lack comprehensive documentation, making it difficult for LLMs to distinguish the core business logic from the auxiliary code~\cite{gang2009business, pan2025code}. This impedes the generation of functionally meaningful abstractions.

To address these challenges, we present \textbf{ArchAgent}, a hybrid agent framework that integrates static analysis, graph-based reasoning, and LLM-powered semantic synthesis. ArchAgent reconstructs accurate and interpretable architectural views from vast, cross-repository codebases. Two corresponding key innovations are introduced. \textbf{1. Scalable architecture understanding.} By combining LLM-based summarization with adaptive grouping, ArchAgent produces semantically enriched code architecture diagrams. File clustering and contextual pruning support scalable input while preserving architectural fidelity. \textbf{2. Business-aligned architecture diagrams.} At the project level, ArchAgent integrates cross-repository information and DevOps-inspired heuristics to identify business-critical modules and generate system-level view architecture diagrams highlights key modules and execution flows aligned with the behavior of real-world systems.

\section{Related Work}\label{sec:related-work}

\begin{figure*}[b]
\centering
\vspace{-0.4cm}
\includegraphics[width=1\linewidth]{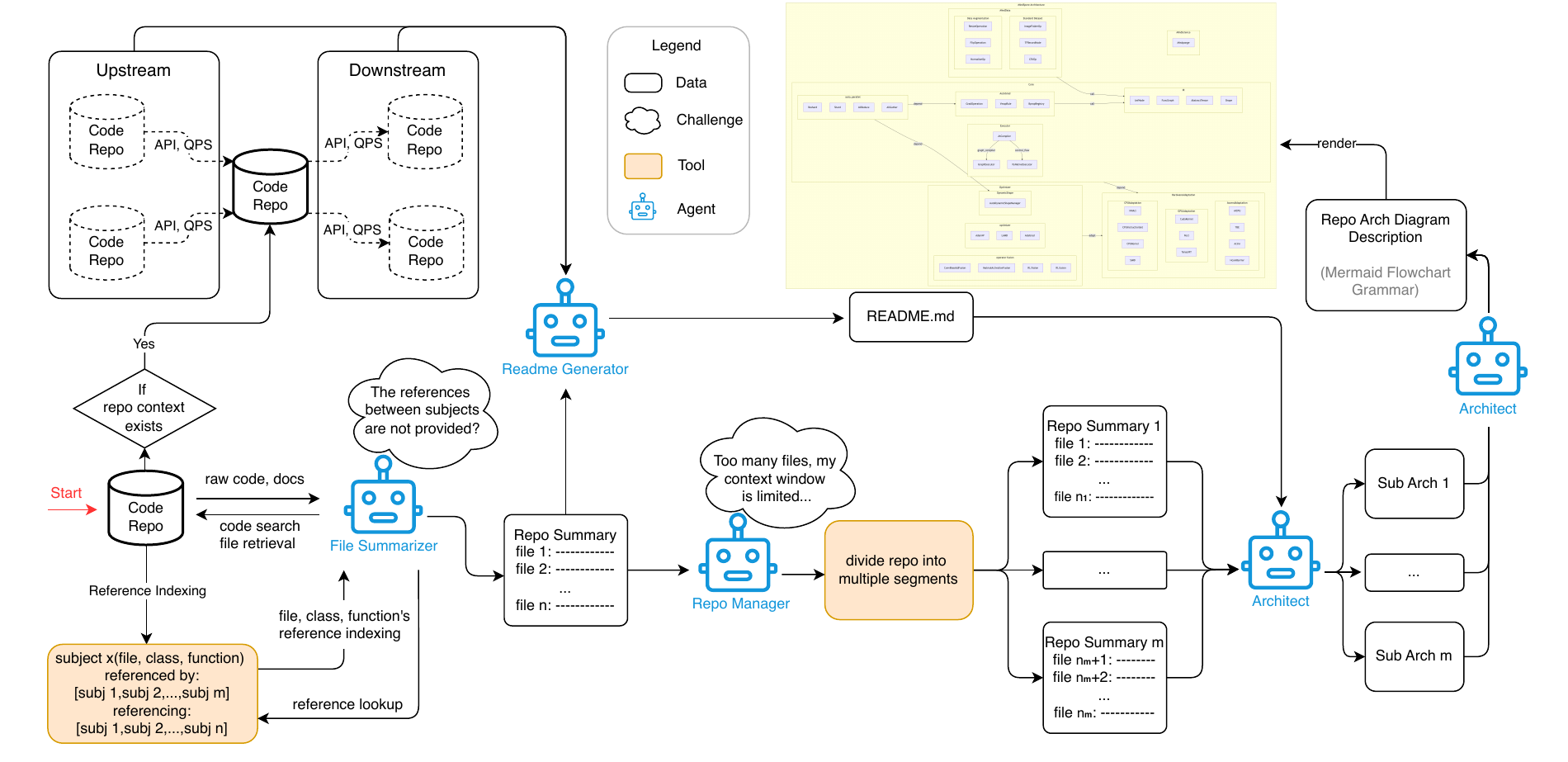}
\vspace{-1.1cm}
\caption{The overall workflow of ArchAgent.}
\label{fig:archagent-workflow}
\end{figure*}

\textbf{Static and Dynamic Architecture Recovery}. Classical techniques extract structural dependencies (calls, imports, inheritance, data flow), and then cluster entities to form architectural components. Representative examples include Algorithm for Comprehension-Driven Clustering (ACDC)~\cite{DBLP:conf/wcre/TzerposH00}, Architectural Recovery using Concerns (ARC)~\cite{DBLP:conf/kbse/GarciaPMMC11}, and Bunch~\cite{DBLP:conf/icsm/MancoridisMCG99}. Despite being effective for small-to-midsize projects, these methods have two major limitations: they are sensitive to input granularity (e.g., results vary between file-level and symbol-level analysis) and they lack semantic context, failing to capture the business logic in code identifiers, comments, and history. This hinders their practical use, where accurate and maintainable architectural views are essential.

\textbf{LLMs for Code Understanding}. LLMs have recently excelled at microlevel tasks such as code completion, defect repair, and single-function summarization~\cite{DBLP:journals/ese/SchneiderBLSBCST25, zhou2025using}.  However, extension to system-level reasoning exposes two fundamental gaps. First, there is a local-scope bias. Since LLMs are often trained in snippet-scale corpora, they may omit functions, misstate control flow, or hallucinate identifiers when asked to summarize whole files or packages~\cite{DBLP:conf/icse/SunMLZFLDLC25}. Second, the finite context window limits their capacity. Even 128k-token models cannot load a million-line repository in one pass, preventing holistic analysis~\cite{DBLP:conf/icse/0009W0LRLW24}. Few prior study has achieved LLM-based end-to-end architecture reconstruction for industrial codebases~\cite{DBLP:journals/corr/SchmidtMC14}. The field therefore lacks a solution that is semantics-aware, business-context-aware, and scalable.

\section{Approach}\label{sec:approach}

ArchAgent analyzes code repositories through a structured pipeline (see \autoref{fig:archagent-workflow}). It begins with raw source code and documentation, potentially involving multiple services. \agent{File Summarizer} performs code search and reference indexing to establish cross-file relationships. \agent{Repo Manager} then segments the codebase into LLM-compatible chunks. \agent{Readme Generator} synthesizes a repository overview using summarized content, API specs, and QPS metrics. Finally, \agent{Architect} generates partial architecture diagrams in Mermaid, which are merged into a complete diagram. Prompts are available online\footnote{https://github.com/panrusheng/ArchAgent-supplement}. Key strategies are detailed below.

\textbf{Adaptive Grouping}. When processing large-scale code repositories, it is crucial to partition the data into manageable groups for efficient input of the model, particularly in LLM-powered code analysis tasks. Traditional fixed-size grouping strategies often result in imbalanced groups, where the last group contains only a few files. Such microgroups can severely impact model understanding due to insufficient context, leading to suboptimal performance in tasks like code summarization or defect detection. To address this issue, we propose an adaptive grouping scheme (used by \agent{Repo Manager} in the workflow) that ensures uniform group sizes by dynamically adjusting the partition based on the token count of the repository summary. Our approach minimizes the variance in group sizes and prevents the formation of undersized tail groups. The core mechanism involves the following steps:

\begin{enumerate}
    \item Calculate total token $ T $ of the whole repository summary.
    \item Define a maximum token threshold $ M $, which represents the upper limit per model input, constrained by the LLM's context window.
    \item Partition the repository into $ G = \lceil T / M \rceil $ groups, where $ \lceil \cdot \rceil $ denotes the ceiling function. This formula ensures that the total tokens are divided into an integer number of groups, each with a size as close to the average.
    \item  To ensure group independence and minimize intergroup correlation loss, we traverse the file tree using Depth First Search (DFS) while maintaining a 10\% overlap rate (determined empirically) between adjacent groups. For instance, the DFS sequence groups $n$ files as: $[f_1, f_n]$, $[f_{0.9n}, f_{1.9n}]$, ..., under the assumption that all file summaries contain an identical number of tokens.
\end{enumerate}

\newcommand{\subsubsubsection}[1]{\paragraph{#1}}

\textbf{File-Level Summarization}. To construct a more comprehensive understanding of the overall semantics of a file, we propose two key enhancements. 
First, we extracted the reference graphs at three levels: \textit{file-level}, \textit{class-level}, and \textit{function-level}, using Abstract Syntax Tree (AST). 
These reference structures serve as auxiliary context for \agent{File Summarizer}.
Second, the summarizer is required to identify other files that are functionally closely related to the target file before generating its summary, allowing the summary of the target file to better reflect interrelated responsibilities and improving functional awareness across the repository.

\begin{algorithm}
\caption{README Generation}\label{algorithm:readme generation}
\SetKwFunction{Gen}{GenerateReadme}
\KwIn{Repository file summaries $\mathcal{S} = \{s_1, s_2, \dots, s_n\}$, related repository information $\mathcal{I}$}
\KwOut{README $\mathcal{R}$}

Identify top-level entry-point files $ \mathcal{E} = \{e_1,e_2, \dots, e_m\} $ from $\mathcal{S}$\;
\ForEach{$e_i \in \mathcal{E}$}{
    Trace downstream call chain or dependency path $\mathcal{T}_i = \{t_{i_1}, t_{i_2}, \dots, t_{i_l}\}$ from $e_i$ and $\mathcal{S}$\;
}
Model content $C = \{\mathcal{T}_1, \mathcal{T}_2, \dots, \mathcal{T}_m\} \cup \mathcal{S}$\;
\If{$\mathcal{I} \neq \emptyset$}{
    Aggregate cross-repository signals: API usage, QPS, repository docs\;
    Incorporate these signals into model context $C$\;
}
$\mathcal{R} \leftarrow$ \Gen{$C$}\;
\Return $\mathcal{R}$\;
\end{algorithm}

\textbf{README Generation}. To address the underrepresentation of program entry points in existing architecture generation tools, we introduce \agent{Readme Generator}, an LLM-assisted process (\autoref{algorithm:readme generation}) that synthesizes a README from file summaries and cross-repository signals. The approach first identifies key entry points and traces their downstream dependencies, then augments this information with inter-repository interaction signals. The organization format of the model content can refer to readme prompt. The resulting README presents a system-level view that highlights architecture, key modules, and execution flows.

\section{Evaluation}\label{sec:evaluation}
We design two quantitative experiments to evaluate our architecture-generating strategy and the added value of cross-repository context, respectively. In addition, we elaborate on a real-world case study to demonstrate the value of our architecture generation technique in production environments. 

\subsection{Scalability Evaluation}\label{sec:comparision-experiment}

As the established industrial best practice, DeepWiki\footnote{https://deepwiki.org} is gaining popularity in software engineering automation. It generates well-structured code documentation and summarizes repository architectures, including large-scale ones ($>4k$files). Alternative tools like gitdiagram often fail beyond 4kfiles by attempting to input entire codebases into LLMs, exceeding typical context windows ($\sim200k$ tokens). Focusing on large repositories, we compare ArchAgent with DeepWiki on codebases with architecture diagram ground truth. We hypothesize that ArchAgent produces more accurate architecture diagrams (via F1 score) than DeepWiki. ArchAgent uses \texttt{Qwen 3-32B-128K}.

\textbf{Dataset}. We curated a benchmark of eight large-scale production-grade GitHub projects, each with an architecture diagram and document, to evaluate our approach. Previously, similar datasets were scarce. We open-sourced this set\footnote{https://github.com/panrusheng/arch-eval-benchmark} to facilitate future research. Various projects contain 1$k$–22$k$ source files in Go, Java, C/C++, Python, and YAML. Public architecture documentation provides reliable ground truth across heterogeneous codebases and paradigms, making the dataset ideal for studying automated architecture generation.

\textbf{Participants}. To prevent systematic bias, we recruited 30 senior software engineers from diverse company departments, each with $\geq$5 years of development experience and independently maintaining at least one complex code repository. They were pre-screened to ensure the ability to analyze sample architecture diagrams. The cohort was divided into two counterbalanced groups using a within-subjects design. Group A (n=15): DeepWiki outputs were first evaluated, followed by ArchAgent; Group B (n=15): ArchAgent first, followed by DeepWiki. This design controlled for order effects (learning/fatigue) with mandatory breaks between sessions.

\textbf{Setup}. Each participant assessed 16 architecture diagrams (8 repositories $\times$ 2 systems) via the following protocol: 
For each model-generated diagram, participants received the architecture document, the generated diagram, and an evaluation table divided into \textit{layers}, \textit{nodes}, and \textit{edges}. Each section listed all generated elements with ``true"/``false" options and allowed adding omitted items. Participants compared the generated and reference diagrams to verify entries and supplement omissions. We aggregated counts of true positives (TP), false positives (FP), and false negatives (FN) per category, then computed precision ($P$), recall ($R$), and F1 score ($F_1$).

\begin{figure}[ht]
\centering
\vspace{-0.3cm}
\includegraphics[width=0.9\linewidth]{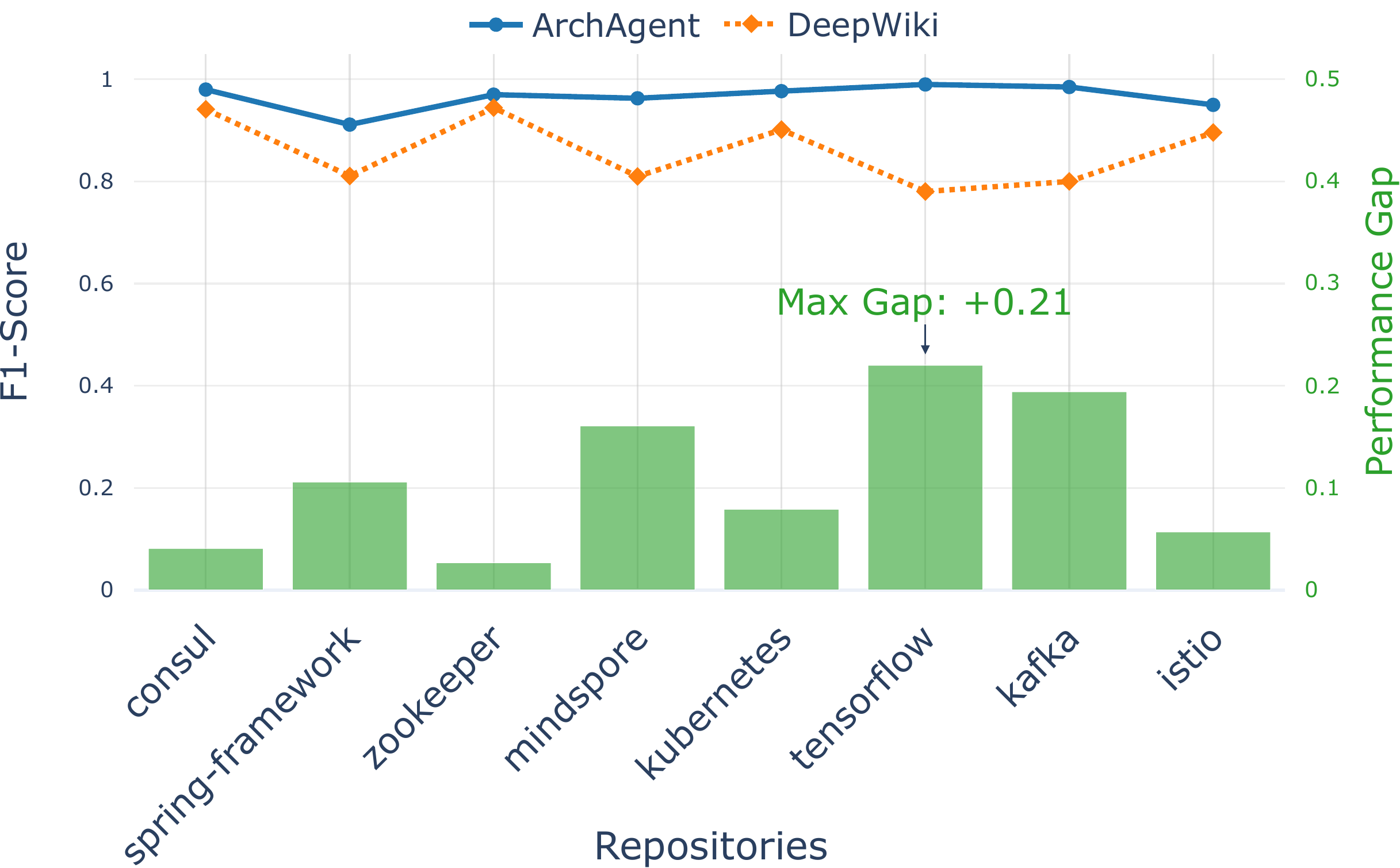}
\vspace{-0.3cm}
\caption{F1-score distribution across repositories: ArchAgent achieves a statistically significant difference versus DeepWiki.}
\vspace{-0.2cm}
\label{fig:f1-github}
\end{figure}

\textbf{Results}. The F1 scores of all repositories are illustrated in \autoref{fig:f1-github}. ArchAgent ($\mu=0.966$, $\sigma=0.025$) achieves a
statistically significant difference compared to DeepWiki ($\mu=0.860$, $\sigma=0.067$), with the p-value of the paired t-test as $0.0036<0.05$ (effect size 1.62, 95\% CI [0.045, 0.165]).

\subsection{Ablation Study on Contextual Dependency}
\label{subsec:ablation}
The central hypothesis posits that integrating repository context enhances architecture diagram precision, especially in reconstructing production repository business logic, with consistent effects across LLMs. To quantify this, a controlled ablation study is designed to isolate the contribution of inter-service dependencies while controlling intra-repository structural information.

\textbf{Dataset \& Participants}. The same 30 participants from the previous experiment (\autoref{sec:comparision-experiment}) also participated in this study. The data set for this experiment was sourced from their industrial real-world repositories, and each participant selected one representative repository they developed (mean size: 425 files, 38,171 lines of code).

\textbf{Procedure}. Two method variants are defined:  
(a) \texttt{Full Method}: generates architectural diagrams incorporating contextual dependencies (intra-repository and inter-service) in README;  
(b) \texttt{Ablated Method}: generates diagrams excluding upstream/downstream dependencies. Each method is run using two open-source models: (a) \texttt{Qwen 3-32B-128K} (Qwen 3); (b) \texttt{DeepSeek-R1-Distill-Llama-70B-64K} (Llama 3). For each repository, four diagram variants are generated (2 methods × 2 models), yielding 120 diagrams total. Each set of four is evaluated by the repository authors. Alongside a pre-defined F1 score, Business Restoration Degree (\(r\)) is introduced, assessed on a 0–100 scale (in 10-point increments) and normalized to \(0-1\). The final score is weighted \(F1 = r \cdot F1\).

\begin{figure}[ht]
\centering
\includegraphics[width=1\linewidth]{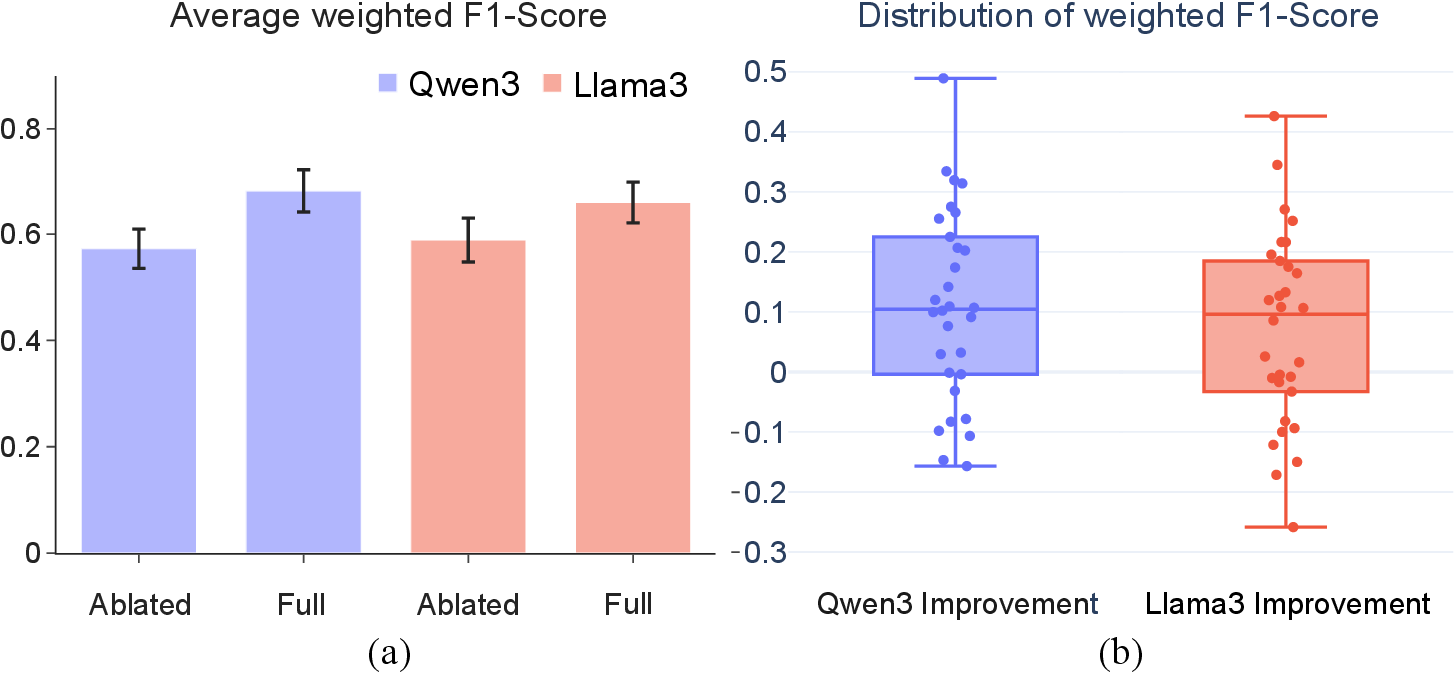}
\vspace{-0.8cm}
\caption{Results of ablation experiment. (a) The mean$\pm$SEM value of four conditions. (b) The distribution of improvements (Full-Albated) on each model.}
\label{fig:f1-intranet}
\end{figure}

\textbf{Results}. Paired t-tests compared \texttt{Full} vs. \texttt{Ablated} performance per model; independent t-tests assessed cross-model differences. As illustrated in \autoref{fig:f1-intranet}(a), the inclusion of dependency contexts yielded significant accuracy improvements. For Qwen 3, contextual F1 increased by 0.11 on average, with a highly
statistically significance ($p=0.00087<0.0001$, effect size 0.68, 95\% CI [0.35, 1.09]). For Llama 3, contextual F1 increased by 0.07 on average with statistically significance ($p=0.023<0.05$, effect size 0.44, 95\% CI [0.12, 0.85]). This confirms our primary hypothesis that dependency modeling is non-redundant for architectural recovery. \autoref{fig:f1-intranet}(b) reveals no significant performance differential between models when using our full method ($p=0.361>0.05$, effect size 0.24, 95\% CI [-0.28, 0.76]), supporting the robustness of our approach across decoder architectures.

\subsection{Case study}
\begin{figure}[ht]
\centering
\includegraphics[width=1\linewidth]{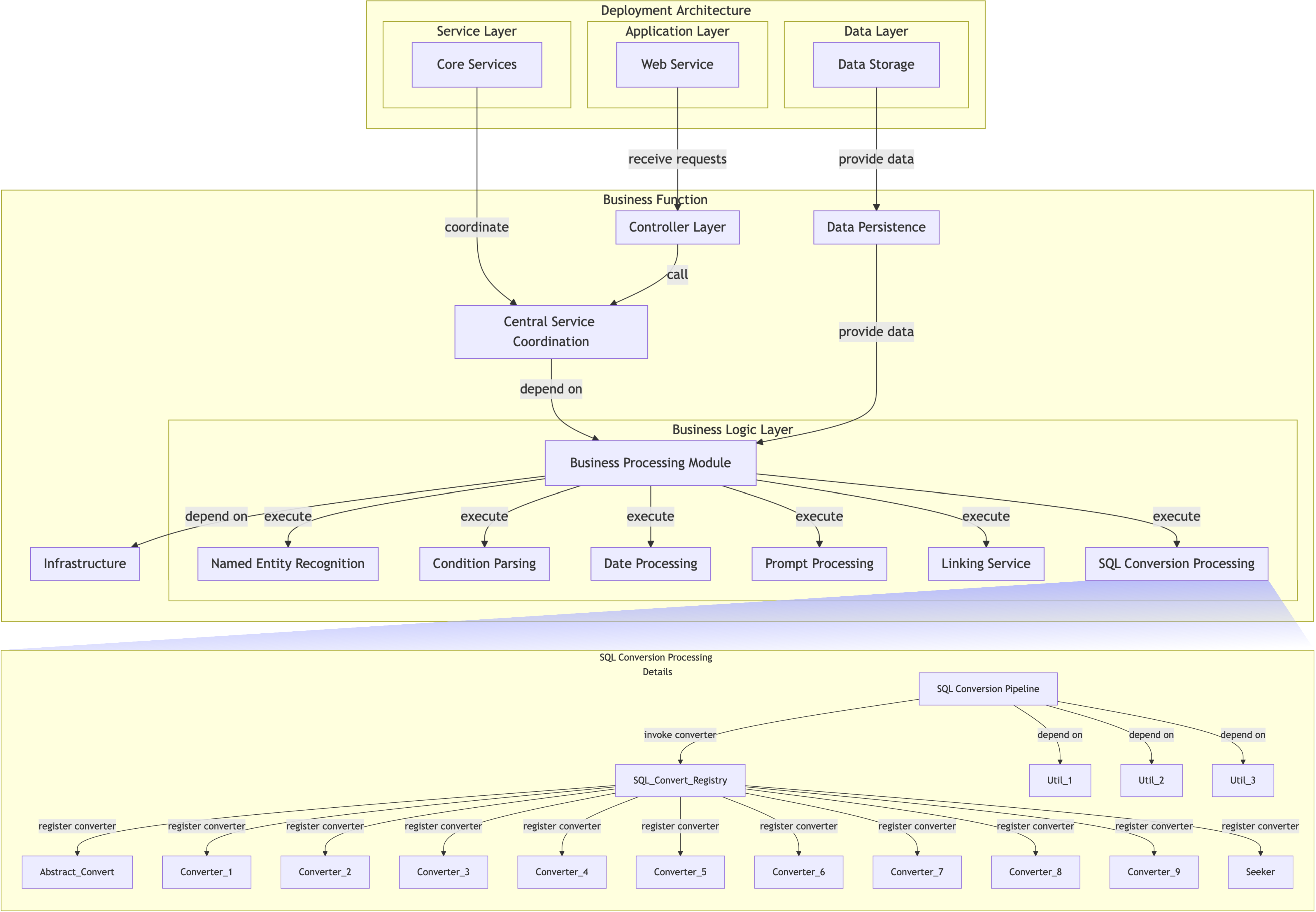}
\vspace{-0.7cm}
\caption{The architecture diagram derived from an industrial-scale SQL transformation codebase.}
\label{fig:case-study}
\end{figure}

From the aforementioned results, we selected a representative repository that implements SQL transformation logic (621 Java files).
The produced diagram \autoref{fig:case-study} automatically surfaced every business–critical module, including parsing, named-entity recognition, control/coordination, and the SQL conversion pipeline. To prevent leakage of the business logic of the repository and ensure the clarity of the diagram, we manually removed all class-level details and replaced every concrete identifier inside the sql-convert module with anonymous numeric labels. It is worth noting that, although the complete SQL conversion process is implemented in a single Java file, our diagram allocates a dedicated, fine-grained subview that visualizes this workflow, thus exposing the core functionality of the repository. The maintainer judged the diagram to provide an almost perfect coverage of the information they usually need to communicate, indicating that our method captures key modules and business concerns effectively. 

\vspace{-0.2cm}
\vspace{-0.6cm}

\section{Conclusion}\label{sec:conclusion}
ArchAgent aims to recover business-aligned architecture from large-scale legacy software systems. By employing adaptive grouping and enhanced summarization strategies, it addresses limitations related to LLM context windows and missing business semantics, facilitating scalable and accurate architecture reconstruction. Evaluations show substantial improvements over industrial baselines and underscore the role of dependency context in capturing business logic.

\bibliographystyle{IEEEbib}
\bibliography{strings,refs}

\begin{thebibliography}{10}

\bibitem{perry1992foundations}
Dewayne~E Perry and Alexander~L Wolf,
\newblock ``Foundations for the study of software architecture,''
\newblock {\em ACM SIGSOFT Software Engineering Notes}, vol. 17, no. 4, pp.
  40--52, 1992.

\bibitem{DBLP:conf/fase/Dayani-FardYMA05}
Homayoun Dayani{-}Fard, Yijun Yu, John Mylopoulos, and Periklis Andritsos,
\newblock ``Improving the build architecture of legacy {C/C++} software
  systems,''
\newblock in {\em Proceedings of the International Conference on Fundamental
  Approaches to Software Engineering}, 2005, vol. 3442, pp. 96--110.

\bibitem{chen2021evaluating}
Mark Chen, Jerry Tworek, Heewoo Jun, Qiming Yuan, Henrique Ponde De~Oliveira
  Pinto, Jared Kaplan, Harri Edwards, Yuri Burda, Nicholas Joseph, Greg
  Brockman, et~al.,
\newblock ``Evaluating large language models trained on code,''
\newblock {\em arXiv preprint arXiv:2107.03374}, 2021.

\bibitem{li2022competition}
Yujia Li, David Choi, Junyoung Chung, Nate Kushman, Julian Schrittwieser,
  R{\'e}mi Leblond, Tom Eccles, James Keeling, Felix Gimeno, Agustin Dal~Lago,
  et~al.,
\newblock ``Competition-level code generation with alphacode,''
\newblock {\em Science}, vol. 378, no. 6624, pp. 1092--1097, 2022.

\bibitem{esposito2025generative}
Matteo Esposito, Xiaozhou Li, Sergio Moreschini, Noman Ahmad, Tomas Cerny,
  Karthik Vaidhyanathan, Valentina Lenarduzzi, and Davide Taibi,
\newblock ``Generative ai for software architecture. applications, trends,
  challenges, and future directions,''
\newblock {\em arXiv preprint arXiv:2503.13310}, 2025.

\bibitem{DBLP:journals/infsof/NguyenVVN25}
Thu{-}Trang Nguyen, Thanh~Trong Vu, Hieu~Dinh Vo, and Son Nguyen,
\newblock ``An empirical study on capability of large language models in
  understanding code semantics,''
\newblock {\em Information and Software Technology}, vol. 185, pp. 107780,
  2025.

\bibitem{dhar2024leveraging}
Rudra Dhar, Karthik Vaidhyanathan, and Vasudeva Varma,
\newblock ``Leveraging generative ai for architecture knowledge management,''
\newblock in {\em Proceedings of the International Conference on Software
  Architecture Companion}, 2024, pp. 163--166.

\bibitem{oskooei2025repository}
Amirkia~Rafiei Oskooei, Selcan Yukcu, Mehmet~Cevheri Bozoglan, and Mehmet~S
  Aktas,
\newblock ``Repository-level code understanding by llms via hierarchical
  summarization: Improving code search and bug localization,''
\newblock in {\em Proceesdings of the International Conference on Computational
  Science and Its Applications}, 2025, pp. 88--105.

\bibitem{li2024retrieval}
Zhuowan Li, Cheng Li, Mingyang Zhang, Qiaozhu Mei, and Michael Bendersky,
\newblock ``Retrieval augmented generation or long-context llms? a
  comprehensive study and hybrid approach,''
\newblock {\em arXiv preprint arXiv:2407.16833}, 2024.

\bibitem{gang2009business}
Xie Gang,
\newblock ``Business rule extraction from legacy system using dependence-cache
  slicing,''
\newblock in {\em Proceedings of the International Conference on Information
  Science and Engineering}, 2009, pp. 4214--4218.

\bibitem{pan2025code}
Zhenyu Pan, Xuefeng Song, Yunkun Wang, Rongyu Cao, Binhua Li, Yongbin Li, and
  Han Liu,
\newblock ``Do code llms understand design patterns?,''
\newblock in {\em Proceedings of the IEEE/ACM International Workshop on Large
  Language Models for Code}, 2025, pp. 209--212.

\bibitem{DBLP:conf/wcre/TzerposH00}
Vassilios Tzerpos and Richard~C. Holt,
\newblock ``{ACDC:} an algorithm for comprehension-driven clustering,''
\newblock in {\em Proceedings of the IEEE Working Conference on Reverse
  Engineering}, 2000, pp. 258--267.

\bibitem{DBLP:conf/kbse/GarciaPMMC11}
Joshua Garcia, Daniel Popescu, Chris Mattmann, Nenad Medvidovic, and Yuanfang
  Cai,
\newblock ``Enhancing architectural recovery using concerns,''
\newblock in {\em Proceedings of the International Conference on Automated
  Software Engineering}, 2011, pp. 552--555.

\bibitem{DBLP:conf/icsm/MancoridisMCG99}
Spiros Mancoridis, Brian~S. Mitchell, Yih{-}Farn Chen, and Emden~R. Gansner,
\newblock ``Bunch: {A} clustering tool for the recovery and maintenance of
  software system structures,''
\newblock in {\em Proceedings of the International Conferenceon Software
  Maintenance}, 1999, p.~50.

\bibitem{DBLP:journals/ese/SchneiderBLSBCST25}
Simon Schneider, Alexander Bakhtin, Xiaozhou Li, Jacopo Soldani, Antonio Brogi,
  Tom{\'{a}}s Cern{\'{y}}, Riccardo Scandariato, and Davide Taibi,
\newblock ``Comparison of static analysis architecture recovery tools for
  microservice applications,''
\newblock {\em Empirical Software Engineering}, vol. 30, no. 5, pp. 128, 2025.

\bibitem{zhou2025using}
Xiyu Zhou, Ruiyin Li, Peng Liang, Beiqi Zhang, Mojtaba Shahin, Zengyang Li, and
  Chen Yang,
\newblock ``Using llms in generating design rationale for software architecture
  decisions,''
\newblock {\em arXiv preprint arXiv:2504.20781}, 2025.

\bibitem{DBLP:conf/icse/SunMLZFLDLC25}
Weisong Sun, Yun Miao, Yuekang Li, Hongyu Zhang, Chunrong Fang, Yi~Liu, Gelei
  Deng, Yang Liu, and Zhenyu Chen,
\newblock ``Source code summarization in the era of large language models,''
\newblock in {\em Proceedings of the International Conference on Software
  Engineering}, 2025, pp. 1882--1894.

\bibitem{DBLP:conf/icse/0009W0LRLW24}
Yuxiao Chen, Jingzheng Wu, Xiang Ling, Changjiang Li, Zhiqing Rui, Tianyue Luo,
  and Yanjun Wu,
\newblock ``When large language models confront repository-level automatic
  program repair: How well they done?,''
\newblock in {\em Proceedings of the International Conference on Software
  Engineering: Companion Proceedings}, 2024, pp. 459--471.

\bibitem{DBLP:journals/corr/SchmidtMC14}
Fr{\'{e}}d{\'{e}}rik Schmidt, Stephen~G. MacDonell, and Andy~M. Connor,
\newblock ``An automatic architecture reconstruction and refactoring
  framework,''
\newblock {\em Computing Research Repository}, vol. abs/1407.6103, 2014.

\end{thebibliography}

\end{document}